\documentclass[showpacs,preprintnumbers,amsmath,amssymb]{revtex4}
\usepackage{bm}
\usepackage{epsfig}
\usepackage{graphicx}
\usepackage{amsmath}
\usepackage{dcolumn}

\begin{document}


\title{Study of the partial wave structure of $\pi^0\eta$
photoproduction on protons}
\author{
A.~Fix$^{1}$\thanks{\emph{eMail address:} fix@tpu.ru},
V.L.~Kashevarov$^{2,3}$\thanks{\emph{eMail address:} kashev@kph.uni-mainz.de},
M.~Ostrick$^{2}$\thanks{\emph{eMail address:} ostrick@kph.uni-mainz.de}}
\affiliation{\mbox{$^1$Tomsk Polytechnic University, Tomsk, Russia}
\\ \mbox{$^2$Institut f\"ur Kernphysik, Johannes Gutenberg-Universit\"at Mainz, Mainz,
Germany} \\
\mbox{$^3$Lebedev Physical Institute, Moscow, Russia}}

\date{\today}

\begin{abstract}
Analysis of the partial wave structure of $\gamma p \to \pi^0 \eta p$ reaction is
presented in the energy region from threshold up to the total center of mass energy $W =
1.9$ GeV. Angular distributions measured with the Crystal Ball/TAPS hermetic detector
system at the Mainz Microtron MAMI are expanded in terms of spherical harmonics. The
relation of the extracted moments to the partial wave structure of the reaction amplitude
is discussed and compared with predictions from model calculations.
\end{abstract}


\maketitle

\section{Introduction}

Presently, we notice enormous world-wide experimental and theoretical efforts to
determine the partial wave structure of photon-induced meson production reactions
unambiguously. This has become possible by combining high-intensity beams and polarized
nucleon targets with hermetic detector systems, a technology which has been available
since a few years at ELSA, JLAB, and MAMI \cite{nstar11}. A unique solution for partial
wave amplitudes is a prerequisite to understand the photoexcitation of baryon resonances
and their decay into $\pi N$, $\eta N$ or $K \Lambda/\Sigma$ final states.

Important complementary information to these efforts is added by the study of
photoproduction of meson pairs like $\pi\pi$ or $\pi\eta$, which allow us to study
resonances which have no significant branching ratio of the decay into the nucleon ground
state. Since the first precise measurements of the $\gamma p\to\pi^0\eta p$ reaction in
Refs.\,\cite{Sendai,GRAAL,Horn,Gutz} the database has been growing significantly
\cite{Kashev,Gutz1} and different models are used in
Refs.\,\cite{Horn,Gutz,Kashev,Gutz1,Doring,FKLO} to interpret the existing data. Some
common results of different model approaches may be summarized as follows. Firstly, the
partial wave with quantum numbers $J^P=3/2^-$ provides the major fraction of the cross
section. It is saturated to a great extent by the $\Delta(1700)3/2^-$ resonance and
perhaps by the $\Delta(1940)3/2^-$ as predicted in Ref.\,\cite{Horn}. The decay of these
baryons into $\eta\Delta$ and finally to $\eta\pi^0 p$ results in the production of
$s$-wave $\eta$ mesons. This is confirmed by the observed distribution of the polar angle
$\theta_\eta$ in the center-of-mass (c.m.) frame which is almost isotropic in the whole
energy region considered. A possible contribution of the $d$-wave to the decay mode
$\Delta(1700)3/2^-\to\eta \Delta$ is small (about 0.1\,$\%$) at the resonance position
$W=1700$ MeV and remains limited with increasing energy. The helicity couplings
$A_{\lambda}$ for the transitions $\gamma p\to \Delta(1700)3/2^-$ with total helicities
$\lambda=1/2$ and $3/2$ are rather close to each other. According to \cite{FKLO} their
ratio varies within a narrow region between 0.95 and 1.3. The role of other resonances,
$\Delta(1600)3/2^+$, $\Delta(1750)1/2^+$, $\Delta(1905)5/2^+$, as well as
$\Delta(1920)3/2^+$, is not significant. They become visible in the angular distribution
mostly due to an interference with the dominant $\Delta3/2^-$ partial wave. In
particular, because of the positive parity, their admixture results in a visible
forward-backward asymmetry in the $\theta_\eta$ distribution. Furthermore, the
contribution of these resonances is important in order to describe the observed circular
photon asymmetries \cite{KashevAphi,FiA11}.

Since the primary goal in the investigation of photoproduction of two pseudoscalars is
the search for nucleon resonances with appreciable two-meson decay modes, special
interest is focused on the partial wave content of these reactions. In such a situation,
model independent analysis of the experiments with polarized particles is of special use.
At the same time, the existing approaches have some weak points which hinder their
effective use as a formal basis for a theoretical analysis of the data. Firstly, the
typical formalism is inherently based on the isobar model, where it is assumed that the
transition to the final $\pi\eta N$ state proceeds via intermediate decay of the
$s$-channel resonances into the quasi-two-body states, containing meson-nucleon isobars,
$P_{33}(1232)$ or $S_{11}(1535)$. In other words, this approach depend on our assumptions
about the production mechanism, so that the corresponding results are to some extent
always model dependent. Secondly, in contrast to single-meson photoproduction, the
conventional scheme for partial wave expansion of the final $\pi\eta N$ state does not
provide multipole representation amenable for practical applications. Indeed, when there
are more than two particles in the final state, angular dependence of the cross section
is governed not only by the total angular momentum, but also by the quantum numbers
related to the two-body subsystems. This makes relation between the spin-parity of the
decaying resonance and the corresponding angular distributions of the final particles
very elaborate and much less transparent in comparison to single meson photoproduction.

In the present paper we would like to demonstrate an alternative method, which is free
from the assumption about dynamics of the resonance decays. The method has already been
used successfully to analyse photoproduction of two neutral pions in
Ref.\,\cite{Kashev2pi0}. The general formalism is described in detail in
Ref.\,\cite{FA2pi0}. Contrary to the usual scheme, where the quantization axis is taken
along the initial beam direction we take it along the normal to the plane spanned by the
moments of the final three particles in the overall center-of-mass frame. The resulting
partial wave expansion of the amplitude is based only on the principles of rotation and
inversion invariance and is therefore completely general.

Since only the unpolarized cross section is considered, we do not try to get detailed
information about the contributing partial waves. Our main object is the moments of
angular distribution of the initial photon beam. Representation of the angular
distribution in terms of the moments may be viewed as an analog to the expansion of the
single-meson photoproduction cross section in terms of Legendre polynomials. In the
latter case the characteristic energy dependence of the expansion coefficients allows one
to gain insight into the partial wave content of the amplitude. In a similar manner in
the present paper we study the partial wave structure of the reaction $\gamma
p\to\pi^0\eta p$ and compare our results with those obtained in earlier studies.

\section{Experimental setup and description of the model} \label{exp_data}

The experimental data for this reaction were obtained recently at the Mainz Microtron
MAMI C. The experiment was performed using the Glasgow-Mainz tagger photon facility and
the Crystal Ball/TAPS spectrometer with a liquid-hydrogen target located in the center of
the Crystal Ball. The tagged photon beam covered the energy range from the reaction
threshold to 1450 MeV with an energy bin about 4 MeV. The spectrometer is a hermetic
system of $NaJ$ and $BaF_2$ crystals covering $97\%$ of the full solid angle. For
charged-particle identification a barrel of 24 scintillation counters surrounding the
target was used. The $\pi^0$ and $\eta$ mesons were identified via their decay into two
photons following $\chi^2$ minimization applied to all possible two-photon combinations.
Parameters of the proton also were measured if it was detectable. The background, which
mainly comes from the $\gamma p\to\pi^0\pi^0 p$ reaction, was subtracted by fitting the
missing-mass distributions for the proton. The systematic uncertainty was estimated to be
5\% and includes uncertainties in the photon flux, target density and detection
efficiency. The experimental setup and the event selection procedure are described in
detail in Ref.\,\cite{Kashev}.

\begin{figure}
\begin{center}
\resizebox{0.5\textwidth}{!}{%
\includegraphics{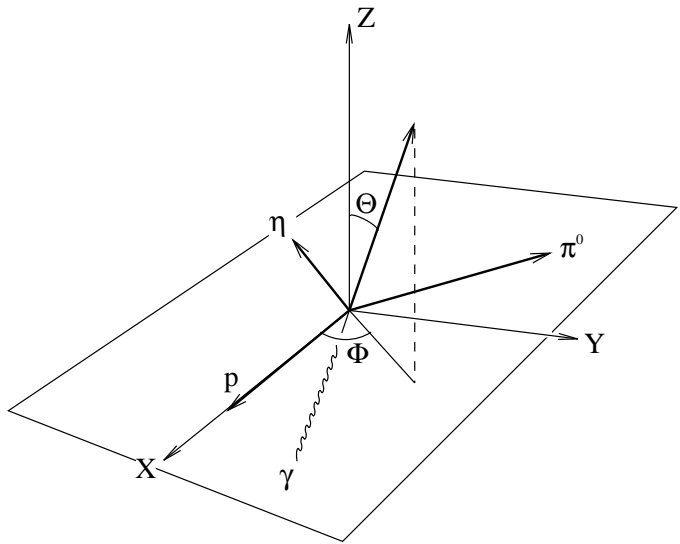}}
\caption{Definition of the coordinate system $XYZ$ in the overall center-of-mass frame.
$\Theta$ and $\Phi$ are respectively the polar and azimuthal angles of the initial
photon.} \label{fig1}
\end{center}
\end{figure}

For the analysis the overall center-of-mass system was used. The coordinate system is
schematically presented in Fig.\,\ref{fig1}. As independent kinematical variables we use
the meson energies $\omega_1$ and $\omega_2$ and the solid angle $\Omega=(\Theta,\Phi)$
fixing the orientation of the initial photon beam with respect to the plane spanned by
the final state particles. The third angle $\Psi$ corresponds to rotation around the
normal. It only influences the cross section in case of linearly polarized particles.

For the $T$-matrix the following expression expanding it over the partial wave amplitudes
was derived in \cite{FA2pi0}
\begin{equation}
T_{\nu\mu}=\sum_{JM_J}t_{\nu\mu}^{JM_J}(W,\omega_1,\omega_2)D_{M_J\mu}^J(\Phi,\Theta,-\Phi)\,.
\end{equation}
Here $D_{M_J\mu}^J$ are the Wigner functions which are matrix elements corresponding to
$2J+1$ representation of the rotation group. For simplicity, the angle $\Psi$ determining
rotation around the normal is taken as zero. The partial wave amplitudes
$t_{\nu\mu}^{JM_J}(W, \omega_1, \omega_2)$ depend on the initial total helicity $\mu$ and
and the final nucleon helicity $\nu$, the total angular momentum $J$, its projection
$M_J$ on the chosen $Z$ axis as well as the total c.m.\ energy $W$ and the meson energies
$\omega_{1}$, $\omega_{2}$.

As in the case of single meson production, a unique determination of the amplitudes
requires much more experimental information about spin-observables than presently is
available. However, using high precision data on the distributions of the normalized
cross sections over the angles $\Theta$ and $\Phi$, it is possible to extract the moments
$W_{LM}$ of an expansion in terms of spherical harmonics
\begin{equation}\label{WThetPhi}
\frac{1}{\sigma}\frac{d\sigma}{d\Omega} =Re\,\sum\limits_{L\geq
0}\sum\limits_{M=-L}^L\sqrt{\frac{2L+1}{4\pi}}\ W_{LM}Y_{LM}(\Theta,\Phi)\,,
\end{equation}
with $W_{00}=1$. These moments are bilinear combinations of the partial wave amplitudes
$t_{\nu\mu}^{JM_J}$ and their measurements put valuable constraints on the partial wave
structure of the full amplitude. The corresponding expression was derived in
Ref.\,\cite{FA2pi0}:
\begin{eqnarray}\label{WLM}
&&W_{LM}=\frac{\pi}{\sigma(2L+1)}{\cal K}\int d\omega_1d\omega_2\sum_{\nu\mu}
\sum_{JJ'M_JM_J'}(-1)^{\mu-M_J} \nonumber\\
&&\phantom{xxxxx}\times C_{J'M_J'J-M_J}^{LM}C_{J'\mu
J-\mu}^{L0}t_{\nu\mu}^{J'M_J'}(W,\omega_1,\omega_2)^*
t_{\nu\mu}^{JM_J}(W,\omega_1,\omega_2)\,,\phantom{xxx}
\end{eqnarray}
where ${\cal K}$ is an appropriate phase space factor and $C_{j_1m_1j_2m_2}^{jm}$ are the
Clebsch-Gordan coefficients for the coupling $\vec{j}_1+\vec{j}_2=\vec{j}$.

As noted in \cite{FA2pi0} the moments $W_{LM}$ obey the following symmetry properties.
Firstly, parity conservation leads to the relation
\begin{equation}
W_{LM}=(-1)^{L+M}W_{LM}\,,
\end{equation}
meaning that the moments with $L+M=$\, odd vanish exactly. Secondly, since
$\frac{1}{\sigma}\frac{d\sigma}{d\Omega}$ in Eq.\,(\ref{WThetPhi}) is a real quantity one
has
\begin{equation}\label{Wconj}
W^*_{LM}=(-1)^{M}W_{L-M}\,,
\end{equation}
leading also to $Im\,W_{L0}=0$. As follows from (\ref{Wconj}), it is sufficient to
consider only the moments with $M\geq0$. Furthermore, waves of the same (opposite) parity
can interfere only for $W_{LM}$ with $L=$\,even (odd). As may readily be seen from
Eq.\,(\ref{WLM}), partial waves of the maximum total angular momentum $J$ can contribute
only to the moments up to $L=2J$. Thus, the amplitudes with $J\leq 3/2$ contribute only
to $W_{11}$ through $W_{3M}$. Using the above properties, a direct determination of the
moments $W_{LM}$ from the experimental data yields valuable information on the states of
total angular momentum that are present in the cross section.

\begin{figure}
\resizebox{0.8\textwidth}{!}{%
\includegraphics{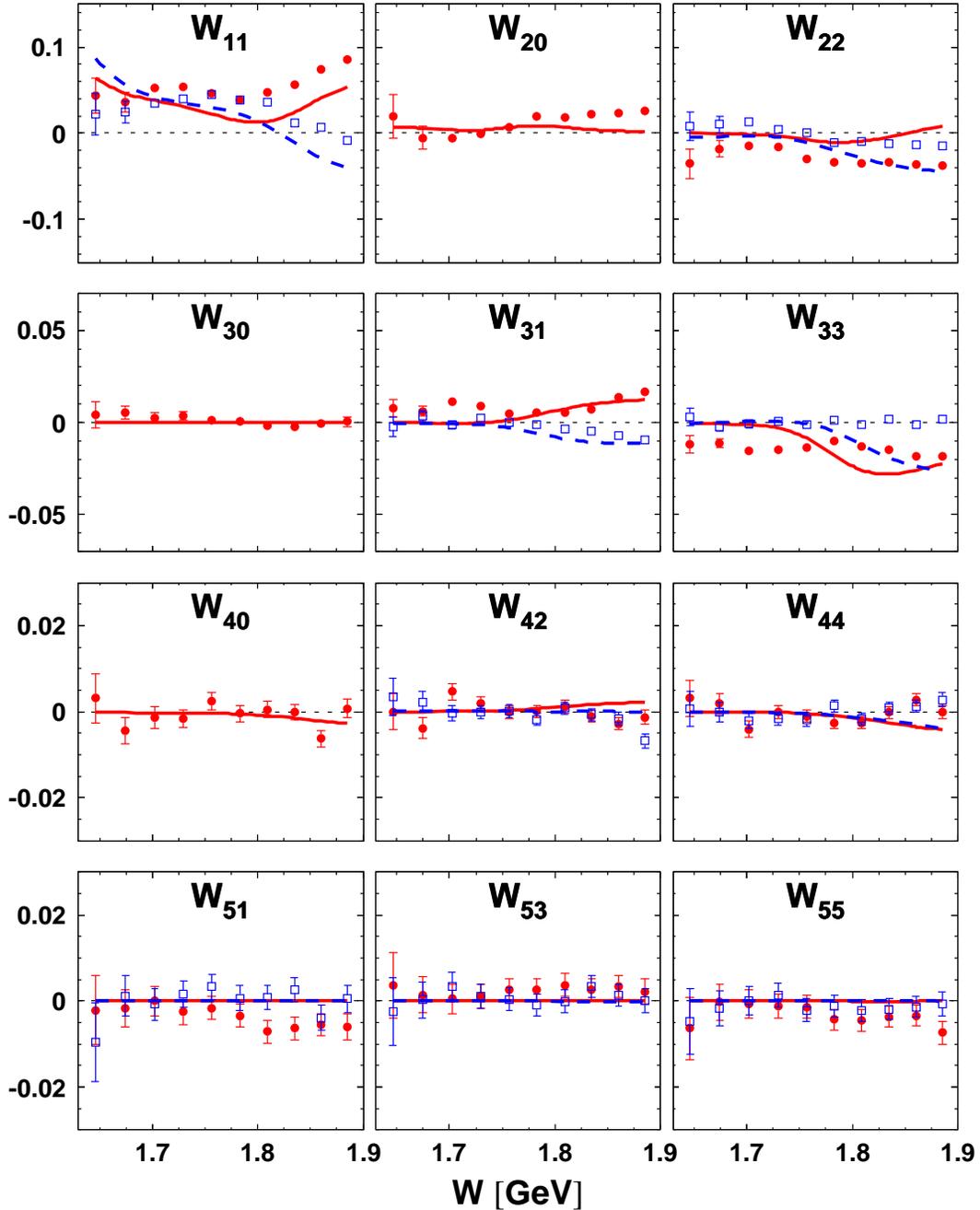}}
\caption{The moments $W_{LM}$ for $\gamma p\to\pi^0\eta p$ as functions of the total
c.m.\ energy $W$, normalized such that $W_{00}=1$. The experimental results for the real
and imaginary parts are shown by filled circles and opened squares, respectively. Only
statistical error bars are shown. The calculation is performed using the isobar model of
Ref.\,\cite{FKLO}. The solid and dashed lines respectively represent the real and
imaginary parts.} \label{fig2}
\end{figure}

\section{Discussion of the results}\label{discussion}

The results for $W_{LM}$ are presented in Fig.\,\ref{fig2} as a function of the total
c.m.\ energy. The moments were obtained from the measured two-dimensional distributions
$\frac{1}{\sigma}\frac{d\sigma}{d\Omega}(\Theta,\Phi)$ using the definition
(\ref{WThetPhi}) as
\begin{equation}
W_{LM}=\frac{1}{\sigma}\,\sqrt{\frac{4\pi}{2L+1}}\int\limits_0^{2\pi}d\Phi\int\limits_0^{2\pi}\frac{d\sigma}{d\Omega}
Y^*_{LM}(\Theta,\Phi)\sin\Theta\,d\Theta.
\end{equation}
The analysis of the results shows that our general notions about the major mechanisms of
the $\gamma p \to \pi^0 \eta p$ reaction based on the previous analyses
\cite{Horn,Doring,FKLO} seem to be correct. Firstly, in the region around $W=1700$ MeV,
because of the closeness of the helicity amplitudes $A_{1/2}$ and $A_{3/2}$ for the
resonance $\Delta(1700)3/2^-$ the moment $W_{20}$ turns out to be small. This result can
easily be explained if one takes into account in the amplitude $t_{\nu\mu}^{\frac32 M_J}$
only the contribution of the resonance $\Delta(1700)3/2^-$. Indeed, from the general
formula (\ref{WLM}) one obtains for $L=2$, $M=0$
\begin{equation}\label{w20}
W_{20}=\frac{\pi}{\sigma}\,{\cal
K}\sum_{\nu}\sum_{M_J=\frac12,\frac32}(-1)^{1/2+M_J}\int\frac{1}{10}\bigg(
\Big|t^{\frac32M_J}_{\nu\,3/2}\Big|^2-\Big|t^{\frac32M_J}_{\nu\,1/2}\Big|^2
\bigg)d\omega_1 d\omega_2\,.
\end{equation}
If only one resonance contributes, the partial wave amplitude $t_{\nu\mu}^{\frac32 M_J}$
can be presented in the factorized form
\begin{equation}
t_{\nu\mu}^{\frac32 M_J}(W,\omega_1,\omega_2)=A_\mu(W)F_\nu^{M_J}(W,\omega_1\omega_2)\,,
\end{equation}
where the vertex function $A_\mu(W)$, $\mu=\pm 1/2,\pm 3/2$, determines electromagnetic
transition $\gamma p\to\Delta(1700)$ with a definite total helicity $\mu$ of the $\gamma
p$ state. It is equal to the corresponding helicity amplitude $A_\mu$ at the resonance
position $W=M_\Delta$. The functions $F_\nu^{M_J}$ which are independent of $\mu$
describe propagation and decay of $\Delta(1700)3/2^-$ into $\pi^0\eta p$. Then for each
value of the $Z$-projection $M_J$, the integrand in Eq.\,(\ref{w20}) is proportional to
the difference $|A_{3/2}|^2-|A_{1/2}|^2$ and, therefore, vanishes in the limit
$A_{3/2}=A_{1/2}$. At the same time, the moment $W_{11}$, which is sensitive to the
interference between $\Delta3/2^-$ and the positive parity states $\Delta1/2^+$ and
$\Delta3/2^-$ is essential. Furthermore, the moments with $L>3$ appearing due to a
possible admixture of the states with higher values of $J$ (e.g., $\Delta5/2^+$) are
expected to be small.

As we can see in Fig.\,\ref{fig2}, the measurements confirm the general smallness of the
moment $W_{20}$ in the region of $\Delta(1700)3/2^-$. The relatively large value of
$W_{11}$ is also in rather well agreement with our assumption about importance of partial
wave amplitudes with positive parity in $\pi^0\eta$ photoproduction. According to the
results of Refs.\,\cite{Horn,FKLO} these partial waves are saturated by the resonances
$\Delta(1750)1/2^+$ and $\Delta(1920)3/2^+$ with a moderate contribution coming from
$\Delta(1600)3/2^+$, $\Delta(1905)5/2^+$, and from non-resonant background. The
insignificance of the moments with $L=3$ shows that a large contribution of the wave
$\Delta3/2^+$ is unlikely. The moments with $L> 3$ are also relatively small. We conclude
that there is little evidence for strong contributions of the waves with total angular
momentum $J>3/2$. In particular the waves $\Delta5/2^-$ and $\Delta5/2^+$ seem not to be
important. As an example of a theoretical description we present in Fig.\,\ref{fig2} the
predictions of the model \cite{FKLO} (solid lines), which are in general agreement with
the data.

\section{Conclusion}\label{conclusion}

We analyzed the moments for the $\gamma p \to \pi^0 \eta p$ reaction using a model
independent expansion of angular distributions of the initial photons in a specific
coordinate system. This allowed us to formulate model independent constraints on the
partial wave structure of the production amplitude. Generally, the data confirm the main
features of the existing model predictions \cite{Horn,Doring,FKLO} for the dynamical
structure of $\pi^0\eta$ photoproduction on a proton. They demonstrate the dominance of
the wave $\Delta3/2^-$, which is expected to come from the resonance $\Delta(1700)3/2^-$
and probably from $\Delta(1940)3/2^-$ as well as the existence of a small admixture of
the resonance states with positive parity. For the higher waves $\Delta5/2^-$ and
$\Delta5/2^+$ no statistically significant evidences were found.


\section*{Acknowledgements}

This work was supported by the Deutsche Forschungsgemeinschaft (SFB 1044). A.F.
acknowledges additional support from the RF Federal programm "Kadry"(contract
14.B37.21.0786), MSE Program 'Nauka' (contract 1.604.2011) and from the Ministry of
education and science of Russian Federation (project 16.740.11.0469).


\end{document}